\documentclass{article}[11pt]
\usepackage{amssymb,amsmath,xspace}
\usepackage[utf8]{inputenc}
\usepackage[T1]{fontenc}
\usepackage[mathscr]{eucal}
\usepackage{color,graphics}
\usepackage[a4paper,margin=2.4cm]{geometry}
\usepackage[pdftex,colorlinks=true,pdfstartview=FitV,
            linkcolor=blue,citecolor=blue,urlcolor=blue]{hyperref}

\hypersetup{
    pdfauthor={Cyril Gavoille, Quentin Godfroy, Laurent Viennot},
    pdftitle={Node-Disjoint Multipath Spanners and their Relationship with Fault-Tolerant Spanners},
    pdfsubject={graph spanners},
    pdfkeywords={graph algorithm,spanner,multipath routing}
}

\usepackage{graphicx}
\usepackage[vlined,english,boxed]{algorithm2e} 

\def\inputfigc#1#2{\begin{center}\scalebox{#1}{\input{#2.pdf_t}}\end{center}}

\def\beginsmall#1{\vspace{-\parskip}\begin{#1}\itemsep-\parskip}
\def\endsmall#1{\end{#1}\vspace{-\parskip}}

\def\le{\leqslant}
\def\ge{\geqslant}
\def\emptyset{\varnothing}

\newtheorem{lemma}{Lemma}

\newtheorem{theorem}{Theorem}
\newtheorem{property}{Property}
\newtheorem{proposition}{Proposition}

\newenvironment{proof}{\noindent\textbf{Proof.}}{{}\hfill$\Box$\\}
\newenvironment{proofof}{\noindent\textbf{Proof of~}}{{}\hfill$\Box$\\}

\newcommand{\pare}[1]{\left({#1}\right)}

\newcommand{\set}[1]{\left\{{#1}\right\}}

\newcommand{\tO}{{\tilde{O}}}

\def\cost{\delta}
\def\bicost{\delta^2}
\def\span{\textbf{A}}
\def\BFS{\mathrm{BFS}}
\def\SPST{{\mathrm{SPST}^2}}

\title{\textbf{Node-Disjoint Multipath Spanners and their Relationship
    with Fault-Tolerant Spanners}}

\author{%
  Cyril Gavoille\thanks{LaBRI, University of Bordeaux, France,
    \texttt{gavoille@labri.fr}. Supported by the European
    project ``EULER'', the ANR project ``ALADDIN'', and the \'equipe-projet INRIA ``C\'EPAGE''.
    Member of the ``Insitut Universitaire de France''.}
\and
Quentin Godfroy\thanks{LaBRI, Universit\'e Bordeaux-I, \texttt{quentin@godfroy.eu}. Supported
                       by the european project ``EULER'', the ANR project ``ALADDIN'', and the
                       \'equipe-projet INRIA ``C\'EPAGE''.}
\and
Laurent Viennot\thanks{INRIA LIAFA, University Paris Diderot, France,
  \texttt{Laurent.Viennot@inria.fr}. Supported by the European project
  ``EULER'', the ANR project ``ALADDIN'', and the \'equipe-projet INRIA ``GANG''.}
}

\begin{document}
\date{}
\maketitle

\begin{abstract}
  Motivated by multipath routing, we introduce a multi-connected
  variant of spanners. For that purpose we introduce the $p$-multipath
  cost between two nodes $u$ and $v$ as the minimum weight of a
  collection of $p$ internally vertex-disjoint paths between $u$ and
  $v$.
  Given a weighted graph $G$, a subgraph $H$ is a $p$-multipath
  $s$-spanner if for all $u,v$, the $p$-multipath cost between $u$ and
  $v$ in $H$ is at most $s$ times the $p$-multipath cost in $G$. The
  $s$ factor is called the stretch.

  Building upon recent results on fault-tolerant spanners, we show how
  to build $p$-multipath spanners of constant stretch and of
  $\tO(n^{1+1/k})$ edges\footnote{Tilde-$O$ notation is similar to
    Big-$O$ up to poly-logarithmic factors in $n$.}, for fixed
  parameters $p$ and $k$, $n$ being the number of nodes of the
  graph. Such spanners can be constructed by a distributed algorithm
  running in $O(k)$ rounds.

  Additionally, we give an improved construction for the case
  $p=k=2$. Our spanner $H$ has $O(n^{3/2})$ edges and the
  $p$-multipath cost in $H$ between any two node is at most twice the
  corresponding one in $G$ plus $O(W)$, $W$ being the maximum edge
  weight.

\end{abstract}

\medskip


\section{Introduction}

It is well-known~\cite{ADDJS93} that, for each integer $k\ge 1$, every
$n$-vertex weighted graph $G$ has a subgraph $H$, called
\emph{spanner}, with $O(n^{1+1/k})$ edges and such that for all pairs
$u,v$ of vertices of $G$, $d_H(u,v) \le (2k-1)\cdot d_G(u,v)$. Here
$d_G(u,v)$ denotes the distance between $u$ and $v$ in $G$, i.e., the
length of a minimum cost path joining $u$ to $v$. In other words,
there is a trade-off between the size of $H$ and its \emph{stretch},
defined here by the factor $2k-1$. Such trade-off has been extensively
used in several contexts. For instance, this can be the first step for
the design of a Distance Oracle, a compact data structure supporting
approximate distance query while using sub-quadratic
space~\cite{TZ05,BGSU08,BK06}. It is also a key ingredient for several
distributed algorithms to quickly compute a sparse skeleton of a
connected graph, namely a connected spanning subgraph with only $O(n)$
edges. This can be done by choosing $k = O(\log n)$. The target
distributed algorithm can then be run on the remaining skeleton
~\cite{BE10}. The skeleton construction can be done in $O(k)$ rounds,
whereas computing a spanning tree requires diameter rounds in
general. We refer the reader to~\cite{Pettie07} for an overview on
graph spanner constructions.

However, it is also proved in~\cite{TZ05} that if $G$ is directed,
then it may have no sub-digraph $H$ having $o(n^2)$ edges and constant
stretch, the stretch being defined analogously by the maximum ratio
between the \emph{one-way} distance from $u$ to $v$ in $H$ and the
one-way distance from $u$ to $v$ in $G$.  Nevertheless, a size/stretch
trade-off exists for the \emph{round-trip} distance, defined as the
sum of a minimum cost of a dipath from $u$ to $v$, and a minimum cost
dipath from $v$ to $u$ (see~\cite{CW00,RTZ08}). Similar trade-offs
exist if we consider the \emph{$p$-edge-disjoint multipath} distance
(in undirected graphs) for each $p\ge 1$, that is the minimum sum of
$p$ edge-disjoint paths joining $u$ and $v$, see~\cite{GGV10}.

\subsection{Trade-offs for non-increasing graph metric}

More generally, we are interested in size/stretch trade-offs for
graphs (or digraphs) for some non-increasing graph metric. A
\emph{non-increasing} graph metric $\delta$ associates with each pair
of vertices $u,v$ some non-negative cost that can only decrease when
adding edges.  In other words, $\delta_G(u,v) \le \delta_H(u,v)$ for
all vertices $u,v$ and spanning subgraphs $H$ of $G$. Moreover, if
$\delta_H(u,v) \le \alpha\cdot\delta_G(u,v) +\beta$, then we say that
$H$ is an $(\alpha,\beta)$-spanner and that its \emph{stretch}
(w.r.t. the graph metric $\delta$) is at most $(\alpha,\beta)$. We
simply say that $H$ is an $\alpha$-spanner if $\beta=0$. The \emph{size}
of a spanner is the number of its edges.

In the previous discussion we saw that every graph or digraph has a
spanner $H$ of size $o(n^2)$ and with bounded stretch for graph metrics
$\delta$ such as round-trip, $p$-edge-disjoint multipath, and the
usual graph distance. However, it does not hold for one-way
distance. A fundamental task is to determine which graph metrics
$\delta$ support such size/stretch trade-off. We observe that the
three former graph metrics cited above have the triangle inequality
property, whereas the one-way metric does not.

This paper deals with the construction of spanners for the
vertex-disjoint multipath metric. A {$p$-multipath} between $u$ and
$v$ is
a subgraph composed of the union of $p$ pairwise internally
vertex-disjoint paths joining $u$ and $v$. The \emph{cost} of a
$p$-multipath between $u$ and $v$ is the sum of the weight of the
edges it contains. Given an undirected positively weighted graph $G$,
define $\cost^p_G(u,v)$ as the minimum cost of a $p$-multipath between
$u$ and $v$ if it exists, and $\infty$ otherwise. A $p$-multipath
$s$-spanner is a spanner $H$ of $G$ with stretch at most $s$
w.r.t. the graph metric $\cost^p$. In other words, for all vertices
$u,v$ of $G$, $\cost^p_H(u,v) \le s\cdot\cost^p_G(u,v)$, or
$\cost^p_H(u,v) \le \alpha\cdot\cost^p_G(u,v) + \beta$ if $s =
(\alpha,\beta)$. It generalizes classical spanners as $d_G(u,v) =
\delta^p_G(u,v)$ for $p=1$.

\subsection{Motivations}

Our interest in the node-disjoint multipath graph metric stems from
the need for multipath routing in networks. Using multiple paths
between a pair of nodes is an obvious way to aggregate
bandwidth. Additionally, a classical approach to quickly overcome link
failures consists in pre-computing fail-over paths which are disjoint
from primary paths~\cite{KKKM07,PSA05,NCD01}.  Multipath routing can
be used for traffic load balancing and for minimizing delays. It has
been extensively studied in ad hoc networks for load balancing,
fault-tolerance, higher aggregate bandwidth, diversity coding,
minimizing energy consumption (see~\cite{MTG03} for a quick overview).
Considering only a subset of links is a practical concern in link
state routing in ad hoc networks~\cite{JV09}. This raises the problem
of computing spanners for the multipath graph metric, a first step
towards constructing compact multipath routing schemes.

\subsection{Our contributions}

Our main contribution is to show that sparse $p$-multipath spanners of
constant stretch do exist for each $p \ge 1$. Moreover, they can be
constructed \emph{locally} in a constant number of rounds. More
precisely, we show that:

\begin{enumerate}

\item Every weighted graph with $n$ vertices has a $p$-multipath $kp
  \cdot O{(1+p/k)}^{2k-1}$-spanner of size $\tO(p^2 \cdot n^{1+1/k})$,
  where $k$ and $p$ are integral parameters $\ge 1$. Moreover, such a
  multipath spanner can be constructed distributively in $O(k)$
  rounds.

\item For $p=k=2$, we improve this construction whose stretch
  is~$18$. Our algorithm provides a $2$-multipath $(2,O(W))$-spanner
  of size $O(n^{3/2})$ where $W$ is the largest edge weight of the
  input graph.

\end{enumerate}

Distributed algorithms are given in the classical $\mathcal{LOCAL}$
model of computations (cf.~\cite{Peleg00b}), a.k.a. the free
model~\cite{Linial92}. In this model nodes operate in synchronous
discrete rounds (nodes are also assumed to wake up simultaneously). At
each round, a node can send and/or receive messages of unbounded
capacity to/from its neighbors and can perform any amount of local
computations. Hence, each round costs one time unit. Also, nodes have
unique identifiers that can be used for breaking symmetry. As long as
we are concerned with running time (number of rounds) and not with the
cost of communication, synchronous and asynchronous message passing
models are equivalent.


\subsection{Overview}

Multipath spanners have some flavors of fault-tolerant spanners,
notion introduced in~\cite{CLPR10a} for general graphs. A subgraph $H$
is an \emph{$r$-fault tolerant $s$-spanner} of $G$ if for any set $F$
of at most $r\ge 0$ faulty vertices, and for any pair $u,v$ of vertices
outside $F$, $d_{H\setminus F}(u,v) \le s \cdot d_{G\setminus
  F}(u,v)$.

\begin{figure}[htb!]
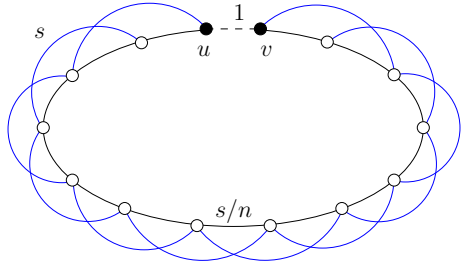

  \inputfigc{.75}{nohop}
  \caption{A weighted graph $G$ composed of a cycle of $n+1$ vertices
    plus $n-1$ extra edges, and a spanner $H =
    G\setminus\set{uv}$. Edge $uv$ has weight~$1$, non-cycle edges
    have weight~$s$, and cycle edges weight~$s/n$ so that $d_H(u,v) =
    s$.  Removing any vertex $z \notin\set{u,v}$ implies
    $d_{G\setminus\set{z}}(u,v) = 1$ and $d_{H\setminus\set{z}}(u,v) =
    2s(1-1/n)$. For other pairs of vertices $x,y$,
    $d_{H\setminus\set{z}}(x,y) / d_{G\setminus\set{z}}(x,y) <
    2s$. Thus, $H$ is a $1$-fault tolerant $2s$-spanner. However
    $\delta^2_H(u,v) / \delta^2_G(u,v) \ge sn/s$. Thus, $H$ is a
    $2$-multipath spanner with stretch at least~$n$.}
  \label{fig:nohop}
\end{figure}

At first glance, $r$-fault tolerant spanners seem related to
$(r+1)$-multipath spanners. (Note that both notions coincide to usual
spanners if $r=0$.) This is motivated by the fact that, if for an edge
$uv$ of $G$ that is not in $H$, and if, for each set $F$ of $r$
vertices, $u$ and $v$ are connected in $H\setminus F$, then by
Menger's Theorem $H$ must contain some $p$-multipath between $u$ and
$v$. If the connectivity condition fulfills, there is no guarantee
however on the cost of the $p$-multipath in $H$ compared to the
optimal one in $G$. Actually, as presented on Fig.~\ref{fig:nohop},
there are $1$-fault tolerant $s$-spanners that are $2$-multipath but
with arbitrarily large stretch.

Nevertheless, a relationship can be established between $p$-mutlipath
spanners and \emph{some} $r$-fault tolerant spanners. In fact, we prove
in Section~\ref{sec:tolerant-to-multipath} that every $r$-fault
tolerant $s$-spanner that is \emph{$b$-hop} is a $(r+1)$-multipath
spanner with stretch bounded by a function of $b,r$ and
$s$. Informally, a $b$-hop spanner $H$ must replace every edge $uv$ of
$G$ not in $H$ by a path simultaneously of low cost and composed of at
most $b$ edges. We observe that many classical spanner constructions
(including the greedy one) do not provide bounded-hop spanners,
although such spanners exist as proved in
Section~\ref{sec:bounded-hop}.  Some variant presented
in~\cite{CLPR10a} of the Thorup-Zwick constructions~\cite{TZ05} are
also bounded-hop (Section~\ref{sec:distributed}).  Combining these
specific spanners with the generic construction of fault tolerant
spanners of~\cite{DK11}, we show in Section~\ref{sec:fault} how to
obtained a $\mathcal{LOCAL}$ distributed algorithm for computing a
$p$-mutlipath spanner of bounded stretch. A maybe surprising fact is
that the number of rounds is independent of $p$ and $n$. We stress
that the distributed algorithm that we obtain has significantly better
running time than the original one presented in~\cite{DK11} that was
$\Omega(p^3\log{n})$.

For instance for $p=2$, our construction can produce a $2$-multipath
$18$-spanner with $O(n^{3/2}\log^{3/2}{n})$ edges. For this particular
case we improve the general construction in Section~\ref{sec:bipath}
with a completely different approach providing a low multiplicative
stretch, namely $2$, at the cost of an additive term depending of the
largest edge weight.

We note that the graph metric $\cost^p$ does not respect the triangle
inequality for $p>1$. For $p=2$, a cycle from $u$ to $w$ and a cycle
from $w$ to $v$ does not imply the existence of a cycle from $u$ to
$v$. The lack of this property introduces many complications for our
second result.  Basically, there are $\Omega(n^2)$ pairs $u,v$ of
vertices, each one possibly defining a minimum cycle $C_{u,v}$ of cost
$\cost^2_G(u,v)$. If we want to create a spanner $H$ with $o(n^2)$
edges, we cannot keep $C_{u,v}$ for all pairs $u,v$. Selecting some
vertex $w$ as pivot for going from $u$ to $v$ is usually a solution of
save edges (in particular at least one between $u$ and $v$). One pivot
can indeed serve for many other pairs. However, without the triangle
inequality, $C_{u,w}$ and $C_{w,v}$ do not give any cost guarantee on
$\cost^2_H(u,v)$.


\section{Main Construction}
\label{sec:first-result}

In this section, we prove the following result:

\begin{theorem}\label{th:first}
  Let $G$ be a weighted graph with $n$ vertices, and $p,k$ be integral
  parameters $\ge 1$. Then, $G$ has a $p$-multipath $kp\cdot
  O{(1+p/k)}^{2k-1}$-spanner of size $O(kp^{2-1/k} {n}^{1+1/k}
  \log^{2-1/k}{n})$ that can be constructed w.h.p. by a randomized
  distributed algorithm in $O(k)$ rounds.
\end{theorem}

Theorem~\ref{th:first} is proved by combining several constructions
presented now.

\subsection{Spanners with few hops}
\label{sec:bounded-hop}

An $s$-spanner $H$ of a weighted graph $G$ is \emph{$b$-hop} if for
every edge $uv$ of $G$, there is a path in $H$ between $u$ and $v$
composed of at most $b$ edges and of cost at most $s\cdot
\omega(uv)$ (where $\omega(uv)$ denotes the cost of edge $uv$). 
An \emph{$s$-hop spanner} is simply an $s$-hop
$s$-spanner.

If $G$ is unweighted (or the edge-cost weights are uniform), the
concepts of $s$-hop spanner and $s$-spanner coincide. However, not all
$s$-spanners are $s$-hop. In particular, the $(2k-1)$-spanners
produced by the greedy\footnote{For each edge $uv$ in non-decreasing
  order of their weights, add it to the spanner if $d_H(u,v) > s\cdot
  d_G(u,v)$.}  algorithm~\cite{ADDJS93} are not.


For instance, consider a weighted cycle of $n+1$ vertices and any
stretch $s$ such that $1 < s < n$.  All edges of the cycle have unit
weight, but one, say the edge $uv$, which has weight $\omega(uv) =
n/s$. Note that $d_G(u,v) = \omega(uv) > 1$. The greedy algorithm adds
the $n$ unit cost edges but the edge $uv$ to $H$ because $d_H(u,v) =
n \le s \cdot \omega(uv)$ (recall that $uv$ is added only if $d_H(u,v)
> s\cdot d_G(u,v)$).  Therefore, $H$ is an $s$-spanner but it is only
an $n$-hop spanner.

However, we have:

\begin{proposition}\label{prop:bounded-hop}
  For each integer $k\ge 1$, every weighted graph with $n$ vertices
  has a $(2k-1)$-hop spanner with less than $n^{1+1/k}$ edges.
\end{proposition}

\begin{proof}
  Consider a weighted graph $G$ with edge-cost function $\omega$. We
  construct the willing spanner $H$ of $G$ thanks to the following
  algorithm which can be seen as the dual of the classical greedy
  algorithm, till a variant of Kruskal's algorithm:

  \beginsmall{itemize}
\item[(1)] Initialize $H$ with $V(H) := V(G)$ and $E(H) := \emptyset$;
\item[(2)] Visit all the edges of $G$ in non-decreasing order of their
  weights, and add the edge $uv$ to $H$ only if every path between $u$
  and $v$ in $H$ has more than $2k-1$ edges.
  \endsmall{itemize}

  Consider an edge $uv$ of $G$. If $uv$ is not in $H$ then there must
  exist a path $P$ in $H$ from $u$ to $v$ such that $P$ has at most
  $2k-1$ edges. We have $d_H(u,v) \le \omega(P)$. Let $e$ be an edge
  of $P$ with maximum weight. We can bound $\omega(P) \le (2k-1) \cdot
  \omega(e)$. Since $e$ has been considered before the edge $uv$,
  $\omega(e) \le \omega(uv)$. It follows that $\omega(P) \le (2k-1)
  \cdot \omega(uv)$, and thus $d_H(u,v) \le (2k-1)\cdot
  \omega(uv)$. Obviously, if $uv$ belongs to $H$, $d_H(u,v) =
  \omega(uv) \le (2k-1) \cdot \omega(uv)$ as well. Therefore, $H$ is
  $(2k-1)$-hop.


  The fact that $H$ is sparse comes from the fact that there is no
  cycle of length $\le 2k$ in $H$: whenever an edge is added to $H$,
  any path linking its endpoints has more than $2k-1$ edges, i.e., at
  least $2k$.

  We observe that $H$ is simple even if $G$ is not.  It has been
  proved in~\cite{AHL02} that every simple $n$-vertex $m$-edge graph
  where every cycle is of length at least $2k+1$ (i.e., of girth at
  least $2k+1$), must verify the Moore bound:
 $$
 n ~\ge~ 1 + d\sum_{i=0}^{k-1} (d-1)^i ~>~ (d-1)^k
 $$
 where $d = 2m/n$ is the average degree of the graph. This implies
 that $m < \frac{1}{2}(n^{1+1/k} + n) < n^{1+1/k}$.

 Therefore, $H$ is a $(2k-1)$-hop spanner with at most $n^{1+1/k}$
 edges.
\end{proof}

\subsection{Distributed bounded hop spanners}
\label{sec:distributed}

There are distributed constructions that provide $s$-hop spanners, at
the cost of a small (poly-logarithmic in $n$) increase of the size of
the spanner compared to Proposition~\ref{prop:bounded-hop}.

If we restrict our attention to deterministic algorithms,
\cite{DGPV08b} provides for unweighted graphs a $(2k-1)$-hop spanner
of size $O(kn^{1+1/k})$. It runs in $3k-2$ rounds without any prior
knowledge on the graph, and optimally in $k$ rounds if $n$ is
available at each vertex.

\begin{proposition}\label{prop:distributed}
  There is a distributed randomized algorithm that, for every
  weighted graph $G$ with $n$ vertices, computes w.h.p. a $(2k-1)$-hop
  spanner of $O(k n^{1+1/k}\log^{1-1/k}{n})$ edges in $O(k)$ rounds.
\end{proposition}

\begin{proof}
  The algorithm is a distributed version of the spanner algorithm used
  in~\cite{CLPR10a}, which is based on the sampling technique
  of~\cite{TZ05}. We make the observation that this algorithm can run
  in $O(k)$ rounds. Let us briefly recall the construction
  of~\cite[p.~3415]{CLPR10a}.

  To each vertex $w$ of $G$ is associated a tree rooted at $w$
  spanning the \emph{cluster} of $w$, a particular subset of vertices
  denoted by $C(w)$. The construction of $C(w)$ is a refinement over
  the one given in~\cite{TZ05}. The main difference is that the clusters'
  depth is no more than $k$ edges. The spanner is composed
  of the union of all such cluster spanning trees. The total number of
  edges is $O(k n^{1+1/k}\log^{1-1/k}{n})$. It is proved
  in~\cite{CLPR10a} that for every edge $uv$ of $G$, there is a
  cluster $C(w)$ containing $u$ and $v$. The path of the tree from $w$
  to one of the end-point has at most $k-1$ edges and cost $\le
  (k-1)\cdot\omega(uv)$, and the path from $w$ to the other end-point
  has at most $k$ edges and cost $\le k\cdot \omega(uv)$. This is
  therefore a $(2k-1)$-hop spanner.

  The random sampling of~\cite{TZ05} can be done without any round of
  communications, each vertex randomly select a level independently of
  the other vertices.  Once the sampling is performed, the clusters
  and the trees can be constructed in $O(k)$ rounds as their the depth
  is at most $k$.
\end{proof}

\subsection{Fault tolerant spanners}
\label{sec:fault}

The algorithm of~\cite{DK11} for constructing fault tolerant spanners
is randomized and generic. It takes as inputs a weighted graph $G$
with $n$ vertices, a parameter $r\ge 0$, and any algorithm $\span$
computing an $s$-spanner of $m(\nu)$ edges for any $\nu$-vertex
subgraph of $G$. With high probability, it constructs for $G$ an
$r$-fault tolerant $s$-spanner of size $O(r^3 \cdot
m(2n/r)\cdot\log{n})$. It works as follows:

\noindent
Set $H := \emptyset$, and repeat independently $O(r^3\log n)$ times:
\beginsmall{itemize}
\item[(1)] Compute a set $S$ of vertices built by selecting each
  vertex with probability $1 - 1/(r+1)$;
\item[(2)] $H := H \cup \span(G\setminus S)$.
\endsmall{itemize}

Then, they show that for every fault set $F \subset V(G)$ of size at
most $r$, and every edge $uv$, there exists with high probability a
set $S$ as computed in Step~(1) for which $u,v \notin S$ and $F
\subseteq S$. As a consequence, routine $\span(G\setminus S)$ provides
a path between $u$ and $v$ in $G\setminus S$ (and thus also in
$G\setminus F$) of cost $\le s \cdot \omega(uv)$. If $uv$ lies on a
shortest path of $G\setminus F$, then this cost is $\le s \cdot
d_{G\setminus F}(u,v)$. From their construction, we have:

\begin{proposition}\label{prop:DK}
  If $\span$ is a distributed algorithm constructing an $s$-hop
  spanner in $t$ rounds, then algorithm~\cite{DK11} provides a
  randomized distributed algorithm that in $t$ rounds constructs
  w.h.p. an $s$-hop $r$-fault tolerant spanner of size $O(r^3\cdot
  m(2n/r)\cdot\log{n})$.
\end{proposition}

\begin{proof}
  The resulting spanner $H$ is $s$-hop since either the edge $uv$ of
  $G$ is also in $H$, or a path between $u$ and $v$ approximating
  $\omega(uv)$ exists in some $s$-hop spanner given by algorithm
  $\span$. This path has no more than $s$ edges and cost $\le
  s\cdot\omega(uv)$.

  Observe that the algorithm~\cite{DK11} consists of running in
  parallel $q = O(r^3\log{n})$ times independent runs of algorithm
  $\span$ on different subgraphs of $G$, each one using $t$
  rounds. Round $i$ of all these $q$ runs can be done into a single
  round of communication, so that the total number of rounds is
  bounded by $t$, not by $q$.

  More precisely, each vertex first selects a $q$-bit vector, each bit
  set with probability $1-1/(r+1)$, its $j$th bit indicating whether
  it participates to the $j$th run of $\span$.  Then, $q$ instances of
  algorithm $\span$ are run in parallel simultaneously by all the
  vertices, and whenever the algorithms perform their $i$th
  communication round, a single message concatenating the $q$ messages
  is sent. Upon reception, a vertex expands the $q$ messages and run
  the $j$th instance of algorithm $\span$ only if the $j$th bit of its
  vector is set.

  The number of rounds is no more than $t$.
\end{proof}

\subsection{From fault tolerant to multipath spanner}
\label{sec:tolerant-to-multipath}

\begin{theorem}\label{th1}
  Let $H$ be a $s$-hop $(p-1)$-fault tolerant spanner of a weighted
  graph $G$. Then, $H$ is also a $p$-multipath $\varphi(s,p)$-spanner
  of $G$ where $\varphi(s,p) = sp \cdot O{(1+p/s)}^s$ and
  $\varphi(3,p) = 9p$.
\end{theorem}

To prove Theorem~\ref{th1}, we need the following intermediate result,
assuming that $H$ and $G$ satisfy the statement of Theorem~\ref{th1}.

\begin{lemma}\label{lem1}
  Let $uv$ be an edge of $G$ of weight $\omega(uv)$ that is not in
  $H$. Then, $H$ contains a $p$-multipath connecting $u$ to $v$ of
  cost at most $\varphi(s,p) \cdot \omega(uv)$ where $\varphi(s,p) = sp \cdot
  O{(1+p/s)}^s$ and $\varphi(3,p) = 9p$.
\end{lemma}

\begin{proof}
  From Menger's Theorem, the number of pairwise vertex-disjoint paths
  between two non-adjacent vertices $x$ and $y$ equals the minimum
  number of vertices whose removal disconnects $x$ and $y$.

  By definition of $H$, $H\setminus F$ contains a path $P_F$ of at
  most $s$ edges between $u$ and $v$ for each set $F$ of at most $p-1$
  vertices (excluding $u$ and $v$). This is because $u$ and $v$ are
  always connected in $G\setminus F$, precisely by a single edge path
  of cost $\omega(uv)$. Consider $P_H$ the subgraph of $H$ composed of
  the union of all such $P_F$ paths (so from $u$ to $v$ in $H\setminus
  F$ -- see Fig.~\ref{fig:kappa5} for an example with $p=2$ and
  $s=5$).

  Vertices $u$ and $v$ are non-adjacent in $P_H$. Thus by Menger's
  Theorem, $P_H$ has to contain a $p$-multipath between $u$ and
  $v$. Ideally, we would like to show that this multipath has low
  cost. Unfortunately, Menger's Theorem cannot help us in this task.

  Let $\kappa_s(u,v)$ be the minimum number of vertices in $P_H$ whose
  deletion destroys all paths of at most $s$ edges between $u$ and
  $v$, and let $\mu_s(u,v)$ denote the maximum number of internally
  vertex-disjoint paths of at most $s$ edges between $u$ and $v$.
  Obviously, $\kappa_s(u,v) \ge \mu_s(u,v)$, and equality holds by
  Menger's Theorem if $s=n-1$. Equality does not hold in general as
  presented in Fig.~\ref{fig:kappa5}. However, equality holds if $s$
  is the minimum number of edges of a path between $u$ and $v$, and
  for $s=2,3,4$ (cf.~\cite{LNLP78}).

  \begin{figure}[htb!]
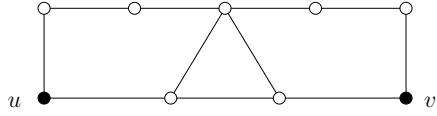

    \inputfigc{.75}{kappa5}
    \caption{A subgraph $P_H$ constructed by adding paths between $u$
      and $v$ with at most $s=5$ edges and with $p=2$. Removing any
      vertex leaves a path of at most~$5$ edges, so $\kappa_5(u,v) >
      1$. However, there aren't two vertex-disjoint paths from $u$ to
      $v$ of at most~$5$ edges, so $\kappa_5(u,v) >
      \mu_5(u,v)$. Observe that $\mu_6(u,v) = \kappa_5(u,v) = 2$.  }
    \label{fig:kappa5}
  \end{figure}

  Since not every path of at most $s$ edges between $u$ and $v$ is
  destroyed after removing $p-1$ vertices in $P_H$, we have that
  $\kappa_s(u,v) \ge p$. Let us bound the total number of edges in a
  $p$-multipath $Q$ of minimum size between $u$ and $v$ in $P_H$. Let
  $r$ be the least number such that $\mu_r(u,v) \ge p$ subject to
  $\kappa_s(u,v)\ge p$. The total number of edges in $Q$ is therefore
  no more than $pr$.


  By construction of $P_H$, each edge of $P_H$ comes from a path in
  $H\setminus F$ of cost $\omega(P_F) \le s\cdot d_{G\setminus F}(u,v)
  \le s\cdot \omega(uv)$. In particular, each edge of $Q$ has weight
  at most $s\cdot \omega(uv)$. Therefore, the cost of $Q$ is
  $\omega(Q) \le prs \cdot \omega(uv)$.

  It has been proved in~\cite{PT93} that $r$ can be upper bounded by a
  function $r(s,p) < \binom{p+s-2}{s-2} + \binom{p+s-3}{s-2} =
  O{(1+p/s)}^s$ for integers $s,p$, and $r(3,p) = 3$ since as seen
  earlier $\kappa_3(u,v) = \mu_3(u,v)$. It follows that $H$ contains a
  $p$-multipath $Q$ between $u$ and $v$ of cost $\omega(Q) \le sp
  \cdot O{(1+p/s)}^s \cdot\omega(uv)$ as claimed.
\end{proof}

\begin{proofof}\textbf{Theorem~\ref{th1}.~}
  Let $x,y$ be any two vertices of a graph $G$ with edge-cost function
  $\omega$. We want to show $\cost^p_H(x,y) \le \varphi(s,p) \cdot
  \cost^p_G(x,y)$. If $\cost^p_G(x,y) = \infty$, then we are done. So,
  assume that $\cost^p_G(x,y) = \omega(P_G)$ for some minimum cost
  $p$-multipath $P_G$ between $x$ and $y$ in $G$. Note that
  $\omega(P_G) = \sum_{uv \in E(P_G)} \omega(uv)$.

  We construct a subgraph $P_H$ between $x$ and $y$ in $H$ by adding:
  (1) all the edges of $P_G$ that are in $H$; and (2) for each edge
  $uv$ of $P_G$ that is not in $H$, the $p$-multipath $Q_{uv}$
  connecting $u$ and $v$ in $H$ as defined by Lemma~\ref{lem1}.

  The cost of $P_H$ is therefore: 
  \begin{eqnarray*}
    \omega(P_H) ~=~ \sum_{uv \in E(P_H)} \omega(uv)
    ~=~ \pare{\sum_{uv \in E(P_G) \cap E(H)} \omega(uv)} + \pare{\sum_{uv \in
      E(P_G)\setminus E(H)} \omega(Q_{uv})}~.
  \end{eqnarray*}
  By Lemma~\ref{lem1}, $\omega(Q_{uv}) \le \varphi(s,p) \cdot \omega(uv)$.
  It follows that:
  \begin{eqnarray*}
    \omega(P_H) &\le& \varphi(s,p) \cdot \sum_{uv \in E(P_G)} \omega(uv) ~=~
    \varphi(s,p)\cdot \omega(P_G) ~=~ \varphi(s,p) \cdot \cost^p_G(x,y)
  \end{eqnarray*}
  as $\varphi(s,p) \ge 1$ and by definition of $P_G$.

  Clearly, all edges of $P_H$ are in $H$.  Let us show now that $P_H$
  contains a $p$-multipath between $x$ and $y$. We first assume $x$
  and $y$ are non-adjacent in $P_H$. By Menger's Theorem applied
  between $x$ and $y$ in $P_H$, if the removal of every set of at most
  $p-1$ vertices in $P_H$ does not disconnect $x$ and $y$, then $P_H$
  has to contain a $p$-multipath between $x$ and $y$.

  Let $S$ be any set of less than $p-1$ faults in $G$. Since $P_G$ is
  a $p$-multipath, $P_G$ contains at least one path between $x$ and
  $y$ avoiding $S$. Let's call this path $Q$. For each edge $uv$ of
  $Q$ not in $H$, $Q_{uv}$ is a $p$-multipath, so it contains one path
  avoiding $S$. Note that $Q_{uv}$ may intersect $Q_{wz}$ for
  different edges $uv$ and $wz$ of $Q$. If it is the case then there is a
  path in $Q_{uv} \cup Q_{wz}$ from $u$ to $z$ (avoiding $v$ and $w$),
  assuming that $u,v,w,z$ are encountered in this order when
  traversing $Q$. Overall there must be a path connecting $x$ to $y$
  and avoiding $S$ in the subgraph $(Q \cap H) \cup \bigcup_{uv \in Q
    \setminus H} Q_{uv}$. By Menger's Theorem, $P_H$ contains a
  $p$-multipath between $x$ and $y$.

  If $x$ and $y$ are adjacent in $P_H$, then we can subdivide the edge
  $xy$ into the edges $xz$ and $zy$ by adding a new vertex $z$. Denote
  by $P'_H$ this new subgraph. Clearly, if $P'_H$ contains a
  $p$-multipath between $x$ and $y$, then $P_H$ too: a path using
  vertex $z$ in $P'_H$ necessarily uses the edges $xz$ and $zy$. Now,
  $P'_H$ contains a $p$-multipath by Menger's Theorem applied on
  $P'_H$ between $x$ and $y$ that are non-adjacent.

  We have therefore constructed a $p$-multipath between $x$ and $y$ in
  $H$ of cost at most $\omega(P_H) \le \varphi(s,p)\cdot \cost^p_G(x,y)$.
  It follows that $\cost^p_H(x,y) \le \varphi(s,p) \cdot \cost^p_G(x,y)$
  as claimed.
\end{proofof}

Theorem~\ref{th:first} is proved by applying Theorem~\ref{th1} to the
construction of Proposition~\ref{prop:DK}, which is based on the
distributed construction of $s$-hop spanners given by
Proposition~\ref{prop:distributed}. Observe that the number of edges
of the spanner is bounded by $O(kp^3 \cdot m(2n/p) \cdot \log{n}) =
O(kp^{2-1/k} n^{1+1/k}\log^{2-1/k}{n})$.

\section{Bi-path Spanners}
\label{sec:bipath}

In this section we concentrate our attention on the case $p=2$, i.e.,
$2$-multipath spanners or \emph{bi-path} spanners for short.  Observe
that for $p=k=2$ the stretch is $\varphi(3,2) = 18$ using our first
construction (cf. Theorems~\ref{th:first} and~\ref{th1}). We provide
in this section the following improvement on the stretch and on the
number of edges.

\begin{theorem}
  \label{th2}
  Every weighted graph with $n$ vertices and maximum edge-weight $W$
  has a $2$-multipath $(2,O(W))$-spanner of size $O(n^{3/2})$ that can
  be constructed in $O(n^4)$ time.
\end{theorem}

While the construction shown earlier was essentially working on edges,
the approach taken here is more global. Moreover, this construction
essentially yields an additive stretch whereas the previous one is
only multiplicative. Note that a $2$-multipath between two nodes $u$
and $v$ corresponds to an elementary cycle. We will thus focus on
cycles in this section.

An algorithm is presented in Section~\ref{sec:construction}. Its
running time and the size of the spanner are analyzed in
Section~\ref{sec:size}, and the stretch in
Section~\ref{sec:stretch}.

\subsection{Construction}
\label{sec:construction}

Classical spanner algorithms combines the use of trees, balls, and
clusters. These standard structures are not suitable to the graph
metric $\delta^2$ since, for instance, two nodes belonging to a ball
centered in a single vertex can be in two different
bi-components\footnote{A short for $2$-vertex-connected components.}
and therefore be at an infinite cost from each other. We will adapt
theses standard notions to structures centered on edges rather than
vertices.

Consider a weighted graph $G$ and with an edge $uv$ that is not a
cut-edge\footnote{A cut-edge is an edge that does not belong to a
  cycle.}. Let us denote by $G[uv]$ the bi-component of $G$ containing
$uv$, and by $\bicost_H(uv,w)$ the minimum cost of a cycle in subgraph
$H$ passing through the edge $uv$ and vertex $w$, if it exists and
$\infty$ otherwise.



We define a \emph{$2$-path spanning tree of root $uv$} as a minimal
subgraph $T$ of $G$ such that every vertex $w$ of $G[uv]$ belongs to a
cycle of $T$ containing $uv$. Such definition is motivated by the
following important property (see Property~\ref{prop:triangle} in
Section~\ref{sec:stretch}): for all vertices $a,b$ in
$G[uv]\setminus\set{u,v}$, $\bicost_G(a,b) \le \bicost_T(uv,a) +
\bicost_T(uv,b)$. This can be seen as a triangle inequality like
property.

If $\bicost_T(uv,w) = \bicost_G(uv,w)$ for every vertex $w$ of
$G[uv]$, $T$ is called a \emph{shortest $2$-path spanning tree}. An
important point, proved in Lemma~\ref{lem:spst} in
Section~\ref{sec:size}, is that such $T$ always exists and contains
$O(\nu)$ edges, $\nu$ being the number of vertices of $G[uv]$.

In the following we denote by $B^2_G(uv,r) = \set{ w : \bicost_G(uv,w)
  \le r}$ and $B_G(u,r) = \set{ w : d_G(u,w) \le r }$ the
\emph{$2$-ball} (resp. \emph{$1$-ball}) of $G$ centered at edge $uv$
(at vertex $u$) and of radius $r$. We denote by $N_G(u)$ the set of
neighbors of $u$ in $G$. We denote by $\BFS(u,r)$ any shortest path
spanning tree of root $u$ and of depth $r$ (not counting the edge
weights). Finally, we denote by $\SPST_G(uv)$ any shortest $2$-path
spanning tree of root $uv$ in $G[uv]$.

The spanner $H$ is constructed with Algorithm~\ref{algo2} from any
weighted graph $G$ having $n$ vertices and maximum edge weight
$W$. Essentially, the main loop of the algorithm selects an edge $uv$
from the current graph lying at the center of a dense bi-component,
adds the spanner $H$ shortest $2$-path spanning tree rooted at $uv$,
and then destroys the neigborhood of $uv$.


\begin{algorithm}[H]
$F := G$, $H := (\emptyset,\emptyset)$\;
\While{$\exists uv \in E(G)$, $|B^2_G(uv,4W) \cap (N_G(u)
\cup N_G(v))| > \sqrt{n}$}{
$H := H \cup \SPST_{F}(uv) \cup \BFS_G(u,2) \cup
\BFS_G(v,2)$\;
$G := G \setminus (B^2_G(uv,4W) \cap (N_G(u) \cup N_G(v)))$
}
$H := H \cup G$
\caption{Construction of $H$.}
\label{algo2}
\end{algorithm}

\subsection{Size analysis}
\label{sec:size}

%
%
%

The proof of the spanner's size is done in two steps, thanks to the
two next lemmas.

First, Lemma~\ref{lem:spst} shows that the while loop does not add too
much edges: a shortest $2$-path spanning tree with linear size
always exists. It is built upon the algorithm of
Suurballe-Tarjan~\cite{ST84} for finding shortest pairs of
edge-disjoint paths in weighted digraphs.


\begin{lemma}\label{lem:spst}
  For every weighted graph $G$ and for every non cut-edge $uv$ of
  $G$, there is a shortest $2$-path spanning tree of root $uv$ having
  $O(\nu)$ edges where $\nu$ is the number of vertices of $G[uv]$. It
  can be computed in time $O(n^2)$ where $n$ is the number of vertices
  of $G$.
\end{lemma}

\begin{proof}
  In the following, we call $X = G[uv]$. $\SPST_X(uv)$ will therefore be
  equal to $\SPST_G(uv)$.
 
  Let $\nu = |V(X)|$ and $\mu = |E(X)|$.

  We first show that we can reduce our problem to finding a one-to-all
  pair of edge-disjoint paths in a directed graph. In other words, let
  $\mathcal{P}$ be a procedure which yields a $2$-(edge-disjoint)-tree
  rooted in a single vertex $w$ in a directed graph $X'$. We show we
  can derive $\mathcal{P'}$ which yields $\SPST_X(uv)$ from
  $\mathcal{P}$.

  First, remark that the problem of finding $\SPST_X(uv)$ is
  equivalent to finding the same structure but rooted in a single
  vertex $w$ where the edge $uv$ is replaced by $uw,wv$, and the
  weights of each edge $uw$ and $wv$ being equal to half of
  $\omega(uv)$.

  We construct $X'$ as follows: each undirected edge is replaced by
  two edges going in opposite direction and of same weight. Then each
  vertex $a$ is replaced by two vertices $a_1$ and $a_2$ where every
  incoming edge arrives at $a_1$ and every leaving edge leaves from
  $a_2$. An edge going from $a_1$ to $a_2$ is finally added.
  Fig.~\ref{reduction_figure} shows what happens to edges of $X$.

\begin{figure}
\centering{\includegraphics[width=130mm]{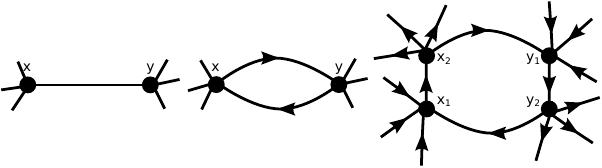}}
\caption{Process by which $X$ is transformed into $X'$.}
\label{reduction_figure}
\end{figure}


Note that $\nu' = |V(X')| = 2 \cdot (\nu + 1)$ and $\mu' = |E(X')| = 2 \cdot
(\mu + 1) + \nu + 1$.

The procedure $\mathcal{P'}$ proceeds as follows:
\begin{enumerate}
\item $uv$ is replaced by $uw,wv$.
\item $X'$ is constructed.
\item $\mathcal{P}$ is called on $X'$, with the root vertex being $w_2$.
\item Every edge of type $x_2 \rightarrow y_1$ present in the result of
$\mathcal{P}$ causes the addition of the edge $xy$ to the result of
$\mathcal{P'}$.
\end{enumerate}

Two edge-disjoint paths in $X'$ are vertex-disjoint in $X$. Indeed, as they
cannot both use an edge of the type $x_1 \rightarrow x_2$ they cannot share
$x_1$ or $x_2$ (except at the extremities) because the only way to leave
(resp. arrive) from $x_1$ (resp. to $x_2$) is to use the edge $x_1
\rightarrow x_2$. So if we have two edge-disjoint paths in $X'$ going from
$w_2$ to some $x_1$, the reduction back to $X$ will yield two disjoint paths
from $w$ to $x$, and then from the edge $uv$ to $x$.

The procedure $\mathcal{P}$ was devised by Suurballe-Tarjan
in~\cite{ST84}.  While not directly constructing the
$2$-(directed-edge-disjoint)-tree rooted in a single vertex $w$
it can be extracted from their algorithm. Roughly speaking, the
construction results of two shortest-path spanning trees computed with
Dijkstra's algorithm. The $2$ paths (from the source to every vertex
$v$) are reconstructed via a specific procedure. This latter can be
analyzed so that the number of edges used in the
$2$-(directed-edge-disjoint)-tree
is at most $2(\nu'-1)$: the structure Suurballe-Tarjan constructed is
such that all vertices, but the source, have two parents.

Therefore the number of edges yielded by procedure $\mathcal{P'}$ is
at most $4 \cdot \nu$, which is $O(\nu)$.
\end{proof}

Secondly, Lemma~\ref{lem:remains} shows that the graph $G$ remaining
after the while loop has only $O(n^{3/2})$ edges. For that, $G$ is
transformed as an unweighted graph (edge weights are set to one) and
we apply Lemma~\ref{lem:remains} with $k=2$. The result we present is
actually more general and interesting in its own right. Indeed, it
gives an alternative proof of the well-known fact that graphs with no
cycles of length $\le 2k$ have $O(n^{1+1/k})$ edges since
$B^2_G(uv,2k) = \emptyset$ in that case.

\begin{lemma}
  \label{lem:remains}
  Let $G$ be an unweighted graph with $n$ vertices, and $k \ge 1$ be
  an integer. If for every edge $uv$ of $G$, $|B^2_G(uv,2k) \cap
  N_G(u)| \le n^{1/k}$, then $G$ has at most $2 \cdot n^{1+1/k}$ edges.
\end{lemma}

\begin{proof}
  Consider Algorithm~\ref{algo_ball} applied to graph $G$.  When the
  procedure terminates, all the vertices and edges of the graph have been
  removed. In the following, we count the number of edges removed by
  each step of the while loop, which in turn allows us to bound the
  number of total edges of $G$.

  \begin{algorithm}
    \For{$i := k-1$ to $0$}{
      \While{$\exists u$, $|B_G(u,i)| \ge n^{i/k}$}
      { $G := G \setminus B_G(u,i)$ } 
    }
    \caption{Remove $1$-balls.}
    \label{algo_ball}
  \end{algorithm}

  Let $X_i$ denote the set of vertices $u$ whose $1$-ball $B_G(u,i)$ is
  removed during iteration $i$ of the {\bf for} loop.  Let $m(u)$ be the
  number of edges erased when removing $B_G(u,i)$. Note that as
  $\sum_i{\sum_{u\in X_i}{|B_G(u,i)|}} = n$ (the procedure removed all
  the vertices), and that $\sum_i{|X_i| \cdot n^{i/k}} \le n$ because
  each $1$-ball is larger than $n^{i/k}$.


  At each step, we argue that
  $$
  m(u) ~\le~ (n^{1/k} + 1) \cdot |B_G(u,i)| + |N_G(u,i+1)|
  $$
  where $N_G(u,i+1)$ is the set of vertices at exactly $i+1$ hops from $u$.

  To this effect, let's consider a shortest path tree $T$ rooted in $u$ and
  spanning $B_G(u,i)$.

  The number of edges in $T$ is bounded by $|B_G(u,i+1)|$, which can be
  decomposed in $|B_G(u,i)| + |N_G(u,i+1)|$.

  We can also bound the total number of non-tree edges as follows: for any
  $x \in B_G(u,i)$, let's consider $B^2_G(xy,2k) \cap N_G(x)$, where $y$ is
  the parent of $x$ in $T$. We know that the number of vertices in this
  $2$-ball is less than $n^{1/k}$ because it is a property of $G$.  But
  $|B^2_G(xy,2k) \cap N_G(x)|$ is also at least the number of non-tree edges
  attached to $x$: for an edge $xz \notin T$, the paths from $z$ towards the
  root $u$ and from $x$ towards the root until they reach common vertex are of
  length at most the radius of $B_G(u,i)$, which is $i \leqslant k$, and so
  there is a cycle of length at most $2k$ using the edges $xz$ and $xy$. So
  the total number of non-tree edges is bounded by $n^{1/k} \cdot |B_G(u,i)|$.



  The termination of the while loop during iteration $i+1$ implies
$$|B_G(u,i+1)| < n^{\frac{i+1}{k}}$$
or equivalently:
$$|N_G(u,i+1)|< n^{\frac{i+1}{k}} - |B_G(u,i)|$$
Therefore we have
$$m(u) < n^{1/k}|B_G(u,i)|+n^{\frac{1+i}{k}}$$
And so
$$m(G) = \sum_{u\in\cup_iX_i}{m(u)} < n^{1/k} \sum_i\sum_{u\in X_i}|B_G(u,i)| + \sum_i {|X_i| \cdot
n^{\frac{i+1}{k}}}$$
and as $\sum_i {|X_i| \cdot n^{i/k}} \le n$, we have
$$|E(G)| \le 2 \cdot n^{1+1/k}~.$$
\end{proof}

Combining these two lemmas we have:
\begin{lemma}\label{lem:size}
  Algorithm~\ref{algo2} creates a spanner of size $O(n^{3/2})$ in time
  $O(n^4)$.
\end{lemma}

\begin{proof}
  Each step of the while loop adds $O(n)$ edges from
  Lemma~\ref{lem:spst}, and as it removes at least $\sqrt{n}$ vertices
  from the graph this can continue at most $\sqrt{n}$ times. In total
  the while loop adds $O(n^{3/2})$ edges to $H$.

  After the while loop, the graph $G$ is left with every
  $B^2_G(uv,4W)\cap(N_G(u)\cup N_G(v))$ smaller than $\sqrt{n}$. If we
  change all edges weights to $1$, it is obvious that every
  $B^2_G(uv,4)\cap(N_G(u)\cup N_G(v))$ is also smaller than
  $\sqrt{n}$. Then as $B^2_G(uv,4)\cap N_G(u)$ is always smaller than
  $B^2_G(uv,4)\cap(N_G(u)\cup N_G(v))$ we can apply
  Lemma~\ref{lem:remains} for $k=2$, and therefore bound the number of
  edges added in the last step of Algorithm~\ref{algo2}.

  The total number of edges of $H$ is $O(n^{3/2})$.


  The costly steps of the algorithm are the search of suitable edges $uv$
  and the cost of construction of $\SPST$.

  The search of suitable edges is bounded by the number of edges as an edge $e$
  which is not suitable can be discarded for the next search: removing edges
  from the graph cannot improve $B^2_G(e,4W)$. Then for starting from one
  extremity of each edge a breadth first search of depth $3$ must be computed,
  keeping only the vertices whose path in the search encounters the other
  extremity. The cost of the search is bounded by the number of edges of $G$. So in
  the end the search costs at most $O(n^4)$.

  The cost of building a $\SPST$ is bounded by the running time of
  \cite{ST84}, which at worst costs $O(n^2)$ (the reduction is essentially
  in $O(m+n)$). Since the loop is executed at most $\sqrt{n}$ times, the
  total cost is $O(n^{7/2})$.

  So the total running time is $O(n^4)$.
\end{proof}

\subsection{Stretch analysis}
\label{sec:stretch}

The proof for the stretch is done as follows: we consider $a,b$ two vertices
such that $\bicost_F(a,b) = \ell$ is finite (if it is infinite there is
nothing to prove). We need to prove that the spanner construction is such
that at the end, $\bicost_H(a,b) \le 2 \ell + O(W)$~. To this effect, we
define $P_F = P^1_F \cup P^2_F$ as a cycle composed of two disjoint paths
($P^1_F$ and $P^2_F$) going from $a$ to $b$ such that its weight sums to
$\bicost_F(a,b)$.

Proving the stretch amounts to show that there exists a cycle $P_H = P^1_H
\cup P^2_H$ joining $a$ and $b$ in the final $H$, with cost at most $2\ell +
O(W)$~. Observe that if the cycle $P_F$ has all its edges in $H$ then one
candidate for $P_H$ is $P_F$ and we are done. If not, then there is at least
one $2$-ball whose deletion provokes actual deletion of edges from $P_F$
(that is edges of $P_F$ missing in the final $H$).

In the following, let $uv$ be the root edge of the first $2$-ball whose
removal deletes edges from $P_F$ (that is they are not added in $H$
neither during the while loop nor the last step of the algorithm). Let $G_i$
be the graph $G$ just before the removal of $B^2_G(uv,4W) \cap (N_G(u) \cup
N_G(v))$~, and $G_{i+1}$ the one just after.

The rest of the discussion is done in $G_i$ otherwise noted.

The proof is done as follows: we first show in Lemma~\ref{lemma_cycle}
that any endpoint of a deleted edge (of $P_F$) belongs to an
elementary cycle comprising the edge $uv$ and of cost at most $6W$. We
then show in Lemma~\ref{simple_path_lemma} that we can construct
cycles using $a$ and/or $b$ passing through the edge $uv$, effectively
bounding $\bicost_H(uv,a)$ and $\bicost_H(uv,b)$ due to the addition
of the shortest $2$-path spanning tree rooted at $uv$. Finally we
show in Lemma~\ref{elem_cycles} that the union of a cycle passing
through $uv$ and $a$ and another one passing through $uv$ and $b$
contains an elementary cycle joining $a$ to $b$, its cost being at
most the sums of the costs of the two original
cycles. 

\begin{lemma}\label{lemma_cycle}
  Let $e = wt$ be an edge of $(G_i \setminus G_{i+1}) \setminus H$.
  Then in $G_i$ both $w$ and $t$ are connected to $uv$ by a cycle of cost at
  most $6W$.
\end{lemma}

\begin{proof}
Since $e$ isn't present neither in $G_{i+1}$ nor $H$, then at least one of
its endpoints is in the vicinity of $u$ or $v$ in $G_i$. W.l.o.g. we can
consider $t$ be a neighbour of $u$ in $G_i$. $w$ is then at most two hops
from $uv$ in $G_i$. The rest of the discussion is done in $G_i$.

We can first eliminate the case where $w$ is a direct
neighbor of $v$, as there is an obvious cycle of $4$ hops: $w \rightarrow t
\rightarrow u \rightarrow v \rightarrow w$~.


Consider now the $\BFS$ tree rooted at $u$ that is added to $H$. As
$w$ is at two hops at most from $u$ there is a path $u\rightarrow
x\rightarrow w$ in this tree ($x$ is defined as the intermediate vertex
of this path and may not exist). As $e$ was removed, it means that
$x$ is distinct from $t$.  Furthermore, $t$ was removed because it
belonged to a $B^2(uv,4W)$, so there is an elementary cycle of at most
$4$ hops passing through $t$ and the edge $uv$.

Now we distinguish two cases as illustrated by
Fig.~\ref{figure_lemma_cycle}.

If $x$ is distinct (which is especially true when it does not exist)
from an intermediate vertex between $v$ and $t$ in the cycle, then we
can extract an elementary cycle of at most $6$ hops passing through
$uv$ and $w$ : $w \rightarrow x \rightarrow u \rightarrow v
\rightarrow \rightarrow t \rightarrow w$.


If $x$ is the same as an intermediate vertex between $v$ and $t$, then the
cycle is $w \rightarrow t \rightarrow u \rightarrow v \rightarrow x
\rightarrow w$.

\begin{figure}
\centering{\includegraphics[width=119mm]{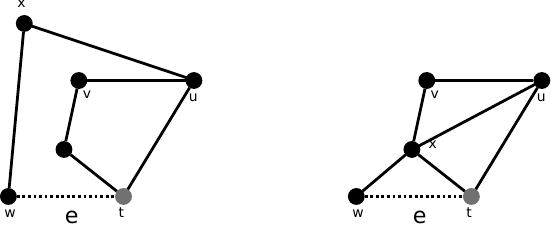}}
\caption{Proof of Lemma~\ref{lemma_cycle}: a cycle of $6$ hops exists
  between $w$ and $uv$ in $G_i$}
\label{figure_lemma_cycle}
\end{figure}
\end{proof}

We now show that we can use this lemma to exhibit cycles going from
$a$ to $uv$ and from $b$ to $uv$.

From the vertices belonging to both $B^2_{G_i}(uv,6W)$ and $P_F$ we choose
the ones which are the closest from $a$ and $b$ (we know that at least two
of them exist because one edge was removed from $P_F$ during step $i$ of the
loop). There are at maximum four of them ($a_1, a_2, b_1, b_2$), one for
each sub-path $P^i_F$ and each extremity $\set{a,b}$. Note that each
extremity is connected to the root edge by an elementary cycle of cost at
most $6W$. Two cases are possible (the placement of the vertices can be seen on
Fig.~\ref{proof_lemma_elementary_cycles}, although the paths on it are from the proof of lemma \ref{elem_cycles}):


\begin{figure}
  \centering{\includegraphics[width=119mm]{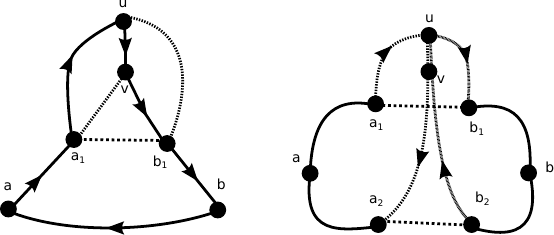}}
  \caption{Proof of Lemma~\ref{elem_cycles}: the two cases for the
    simple paths.} 
  \label{proof_lemma_elementary_cycles}
\end{figure}

\begin{description}

\item[Case 1:] There are only two extremities (then they belong to the
  same subpath) and their cycles which connect them to $uv$ do not
  intersect the second subpath (w.l.o.g we can suppose it is $a_1$ and
  $b_1$).

\item[Case 2:] There are more than two extremities: either some edges
  of the second path were removed or one of the cycles going from one
  of the extremities $a_1$ or $b_1$ to $uv$ intersects the second
  path. 

\end{description}


We show next that we can bound $\bicost_H(uv,a)$ and $\bicost_H(uv,b)$
with the help of the cycles connecting the endpoints and the path
$P_G$. This is done with the two next lemmas.

\begin{lemma}\label{simple_path_lemma}
  For any two vertices joined to the same edge $uv$ by elementary cycles
  there is a simple path of cost at most the sum of the cycles' costs
  and passing through the edge $uv$.
\end{lemma}

\begin{proof}
  Let us call $w$ and $t$ the two vertices. We will show there is a
  simple path going from $w$ to $t$ passing through $uv$. Let us call
  $Q^1$ the elementary cycle joining $w$ to $uv$ and $Q^2$ the one
  joining $t$ to $uv$.  If by following $Q^1$ to reach from $w$ one of
  the endpoints of $uv$ we do not encounter $Q^2$, then the path from
  $w$ to $t$ is composed of the part of $Q^1$, then the edge $uv$,
  then the part of $Q^2$ which reaches $t$ without passing through
  $uv$. If it is not possible, then there are intersection points
  between $Q^1$ and $Q^2$. Let $i$ be the closest intersection point
  from $w$. The path we are looking for is therefore $w \rightarrow i$
  using $Q^1$ then $i \rightarrow uv \rightarrow t$ using the part of
  $Q^2$ which uses the edge $uv$ (the other part would take us
  directly to $a_2$ whithout using $uv$). This path is simple because
  $Q^2$ is an elementary cycle and it cannot cross $Q^1$ before $i$
  because of the way $i$ is chosen.  The two cases are shown on
  figure~\ref{figure_proof_lemma_simple_path}.

  \begin{figure}
    \centering{\includegraphics[width=119mm]{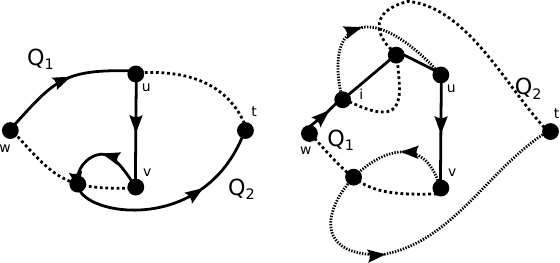}}
    \caption{The two cases for $Q_1$ and $Q_2$. The big dots represent the
      paths' unused portions.}
    \label{figure_proof_lemma_simple_path}
  \end{figure}
\end{proof}


\begin{lemma}\label{elem_cycles}
  Let $a,b$ be two vertices such that an elementary cycle of cost
  $\bicost(a,b)$ has common vertices with some $B^2(uv,6W)$. Then
  $\bicost(a,uv)$ and $\bicost(b,uv)$ are bounded by $\bicost(a,b) + 12W$
\end{lemma}

\begin{proof}
  The lemma is independent of the graph, but for clarity it will be
  proved using the graph $G_i$ and $P_F$.

  Recall that we distinguished two cases depending on whether
  $B^2_{G_i}(uv,6W)$ intersects only one path of $P_F$ (either ${P^1}_F$
  or ${P^2}_F$) or both. Fig.~\ref{proof_lemma_elementary_cycles}
  illustrates the proof of the two cases.

  Lemma~\ref{simple_path_lemma} allows us to solve the first case :
  since there are no intersection on the second path, the cycle $a
  \rightarrow a_1 \rightarrow uv \rightarrow b_1 \rightarrow b
  \rightarrow a$ is simple.
  So there is a cycle in $G_{i}$ joining $uv$, $a$ and $b$ of cost at
  most $12W+\bicost_{G_{i}}(a,b)$. So $\bicost_{G_i}(uv,a)$ is bounded by
  $12W+\bicost_{G_i}(a,b)$ and so is $\bicost_{G_i}(uv,b)$.

  In the second case there are three or four extremities: $a_1$,
  $b_1$, $a_2$ and $b_2$, with possibly $a_2$ and $b_2$ being the same
  vertex. We can apply Lemma~\ref{simple_path_lemma} twice: once between
  $a_1$ and $a_2$ and another time between $b_1$ and $b_2$. These
  create a simple cycle from $a$ to $uv$ passing by $a_1$ and $a_2$
  and another one from $b$ to $uv$ passing by $b_1$ and $b_2$. We know
  the cycles are simple because the vertices were chosen to be the
  closest from $a$ or $b$.  Note that $a_2$ and $b_2$ can be the
  same. So $$
  \bicost_{G_i}(a,uv) \leqslant \omega(a\rightarrow a_1) + 12W +
  \omega(a_2\rightarrow a) \leqslant \bicost(a,b) + 12W
  $$ and the same for $b$.

\end{proof}




\begin{property}\label{prop:triangle}
  Let $uv$ be a non cut-edge of $G$ and $T$ be any $2$-path
  spanning tree rooted at $uv$. Then, for all vertices $a,b$ in
  $G[uv]\setminus\set{u,v}$, $\bicost_G(a,b) \le \bicost_T(uv,a) +
  \bicost_T(uv,b) - \omega(uv)$.
\end{property}

\begin{proof}
  There is in $T$ a cycle joining $a$ to $uv$ of cost
  $\bicost_T(uv,a)$, and another one joining $b$ to $uv$ of cost
  $\bicost_T(uv,b)$. Consider the subgraph $P$ containing only the
  edges from these two cycles. The cost of $P$ is $\omega(P) \le
  \bicost_T(uv,a) + \bicost_T(uv,b) - \omega(uv)$ as edge $uv$ is
  counted twice. It remains to show that $P$ contains an elementary
  cycle between $a$ and $b$. Note that since $a \notin\set{u,v}$, $a$
  has in $P$ two vertex-disjoint paths leaving $a$ and excluding edge
  $uv$: one is going to $u$, and one to $v$. Similarly for vertex $b$.

  W.l.o.g. we can assume that $a$ and $b$ are not adjacent in
  $P$. Otherwise we can subdivide edge $ab$ to obtain a new subgraph
  $P'$. Clearly, if $P'$ contains an elementary cycle between $a$ and
  $b$, then $P$ too. Consider that one vertex $z$, outside $a$ and
  $b$, is removed in $P$.  From the remark above, in
  $P\setminus\set{z}$, there must exists a path leaving $a$ and
  joining some vertex $w_a \in \set{u,v}\setminus\set{z}$ and one path
  leaving $b$ and joining some vertex $w_b \in
  \set{u,v}\setminus\set{z}$. If $w_a = w_b$, then $a$ and $b$ are
  connected in $P\setminus\set{z}$. If $w_a \neq w_b$, then edge $uv$
  belongs to $P\setminus\set{z}$ since in this case $z
  \notin\set{u,v}$, and thus a path connected $a$ to $b$ in
  $P\setminus\set{z}$. By Menger's Theorem, $P$ contains a
  $2$-multipath between $a$ and $b$.
\end{proof}

\begin{lemma}
  $H$ is a $2$-multipath $(2,24W)$-spanner.
\end{lemma}

\begin{proof}
If there is in $F$ a path of cost $\bicost(a,b)$ such that every edge of it
is in $H$, then there is nothing to prove. If there is some removed edge,
then we can identify the loop order $i$ which removed the first edge, and we
can associate the graph $G_i$ just before the deletion performed in the
second step of the loop (so $P_F$ still completely exist in $G_i$). By
virtue of Lemma~\ref{lemma_cycle} we can identify some root-edge $uv$ and
we know that there are some vertices of $P_F$ linked to $uv$ by an
elementary cycle of length at most $6W$. Lemma~\ref{elem_cycles} can then be
applied, and so in $G_i$, $\bicost_{G_i}(a,uv)$ and $\bicost_{G_i}(b,uv)$
are both bounded by $\bicost_{G_i}(a,b)+12W$. As the loop's first step is to
build a shortest $2$-path spanning tree rooted in $uv$ we know that in $H$
$$
\bicost_H(a,uv) \leqslant \bicost_{G_i}(a,uv) \leqslant
\bicost_{G_i}(a,b)+12W
$$
and the same for $b$. Property~\ref{prop:triangle} can then be used in the
$2$-path spanning tree, to bound $\bicost_H(a,b)$:
$$
\bicost_H(a,b) \leqslant \bicost_H(a,uv) + \bicost_H(b,uv) \leqslant
2\cdot \bicost_{G_i}(a,b) + 24W
$$ Finally, as in $G_i$ $P_F$ still exists completely, we have that
$\bicost_{G_i}(a,b) = \bicost_F(a,b)$, so
$$
\bicost_H(a,b) \leqslant 2 \cdot \bicost_F(a,b) + 24W
$$


\end{proof}


\section{Conclusion}

We have introduced a natural generalization of spanner, the
vertex-disjoint path spanners.  We proved that there exists for
multipath spanners a size-stretch trade-off similar to classical
spanners.  We also have presented a $O(k)$ round distributed algorithm
to construct $p$-multipath $kp \cdot O{(1+p/k)}^{2k-1}$-spanners of
size $\tO(p^2 {n}^{1+1/k})$, showing that the problem is \emph{local}:
it does not require communication between distant vertices.


Our construction is based on fault tolerant spanner. An interesting
question is to know if better construction (in term of stretch) exists
as suggested by our alternative construction for $p=2$.

The most challenging question is to explicitly construct the $p$
vertex-disjoint paths in the $p$-multipath spanner. This is probably
as hard as constructing efficient routing algorithm from sparse
spanner. We stress that there is a significant difference between
proving the existence of short routes in a graph (or subgraph), and
constructing and explicitly describing such short routes. For instance
it is known (see~\cite{GS11}) that sparse spanners may exist whereas
routing in the spanner can be difficult (in term of space memory and
stretch of the routes).


\newpage

\bibliographystyle{alpha}
\bibliography{biblio}


\end{document}